\documentclass[10pt, journal, twocolumn]{IEEEtran} 
\IEEEoverridecommandlockouts

\usepackage[T1]{fontenc}
\usepackage{cite}
\usepackage{amsmath,amssymb}
\usepackage{mathtools}
\usepackage{graphicx}
\usepackage{booktabs}
\usepackage{tabularx}
\usepackage{acronym}
\usepackage{tikz}
\usepackage{algorithm}
\usepackage{float}
\usepackage{algpseudocode}
\usepackage[hidelinks]{hyperref}
\usepackage{verbatim}
\usetikzlibrary{fit,backgrounds,arrows.meta}
\usetikzlibrary{arrows.meta,positioning}

\begin{document}
\bstctlcite{BSTcontrol}

\acrodef{3gpp}[3GPP]{Third Generation Partnership Project}
\acrodef{aiot}[A-IoT]{ambient Internet of Things}
\acrodef{cbra}[CBRA]{contention-based random access}
\acrodef{mac}[MAC]{medium access control}
\acrodef{r2d}[R2D]{reader-to-device}
\acrodef{d2r}[D2R]{device-to-reader}
\acrodef{acgi}[AcGI]{access-guidance information}
\acrodef{ao}[AO]{access occasion}
\acrodefplural{ao}[AOs]{access occasions}
\acrodef{po}[PO]{paging opportunity}
\acrodefplural{po}[POs]{paging opportunities}
\acrodef{pdu}[PDU]{protocol data unit}
\acrodef{iot}[IoT]{Internet of Things}
\acrodef{nack}[NACK]{negative acknowledgement}
\acrodef{ai}[AI]{Artificial Intelligence}

\acrodef{acgi}[AcGI]{access-guidance information}
\title{Adaptive Guidance for 3GPP Ambient IoT Contention-Based Random Access}
\author{\IEEEauthorblockN{
        Amirhossein  Azarbahram,~\IEEEmembership{Member,~IEEE},
        David Ernesto Ruiz-Guirola,~\IEEEmembership{Member,~IEEE},\\
        and Onel L. A. L\'opez,~\IEEEmembership{Senior Member,~IEEE}}
\thanks{This work is supported by the Research Council of Finland (369116 (6G Flagship), and 362782 (ECO-LITE)); and by the European Commission through the Horizon Europe/JU SNS project AMBIENT-6G (Grant 101192113).}
\thanks{Authors are with the Centre for Wireless Communications (CWC), University of Oulu, Finland (e-mails: \{amirhossein.azarbahram, david.ruizguirola, onel.alcarazlopez\}@oulu.fi).}}
\maketitle

\begin{abstract}

The emerging ambient Internet of Things (A-IoT)
envisions connectivity among large populations
of low-cost, energy-constrained devices.
Supporting such connectivity requires efficient
coordination of device access, particularly in
reader-triggered operation, where multiple
devices may respond to paging over limited shared
radio resources. Indeed, simultaneous responses can
lead to severe contention, delayed transmissions,
and lost reports. This article explores how
compact extensions to the current 3GPP Release-19
signaling framework could enable more selective
and adaptive device access. The proposed approach
allows readers to broadcast compact access guidance,
which devices interpret according to their local
conditions, while optional feedback supports
continuous adaptation. Through a device-reporting
case study, we demonstrate how such coordination
can improve report delivery and timeliness under
congested conditions, while highlighting the
trade-offs across different network configurations.
We also examine practical signaling considerations
and discuss opportunities for more intelligent,
adaptive, and resilient protocols.
\end{abstract}

\acresetall

\section{Introduction}\label{sec:introduction}

\Ac{aiot} enables connectivity for low-complexity,
ultra-low-power devices operating with harvested energy
and limited energy storage. By minimizing device
complexity and maintenance requirements, \ac{aiot}
supports large-scale applications such as asset tracking,
inventory management, and environmental monitoring.
To support such applications, the \ac{3gpp} Release-19
framework establishes the architecture and radio-access
procedures for \ac{aiot}~\cite{shen2025overview,3gpp_23369}.

A key challenge in scaling \ac{aiot} is coordinating
access among large numbers of devices sharing
limited radio resources. In the Release-19
reader-triggered procedure~\cite{3gpp_38391}, a paging message
can address individual devices, groups, or the
entire device population. Under \ac{cbra},
paged devices select among the configured
\acp{ao}, potentially resulting in collisions
and unsuccessful access when many devices
respond simultaneously.
Managing such contention is particularly
challenging when the reader has limited
knowledge of pending device reports and
access demand varies across paging occasions.

Managing contention in random-access networks
has long relied on two complementary approaches: i)
adjusting the available access resources and
ii) regulating device participation. Classical
framed-ALOHA schemes adapt contention resources
based on observed outcomes~\cite{schoute1983dynamic},
while access barring and backoff mechanisms
control simultaneous access attempts, often
supported by congestion estimation~\cite{song2025twolevel}.
These principles also extend to \ac{aiot},
where adaptive multi-channel access and
reader-broadcast access probabilities have been
investigated to manage contention among large
device populations~\cite{song2026dynamic,ruizguirola2026adaptive}.
Despite their different implementations,
these approaches primarily regulate access
through shared resource configurations or
participation probabilities.


Beyond regulating aggregate contention, access
decisions can also account for the individual
conditions and requirements of devices.
For instance, age-dependent access schemes
consider information freshness when determining
transmission opportunities~\cite{chen2020age}.
More broadly, context-aware protocol design has
been recognized as an important enabler of
dependable low-power IoT operation, leveraging
information freshness, energy availability,
task relevance, and communication conditions
~\cite{David_context}.
Related ideas appear in \ac{aiot} patent disclosures,
including mechanisms for suppressing repeated
group-paging responses~\cite{patent_group_suppression},
enabling selective responses through paging
thresholds~\cite{patent_paging_threshold}, and
incorporating device energy states and access
history into contention control
~\cite{patent_congestion_control}.


These developments demonstrate the potential of
reader-assisted contention management and
selective device responses. However, existing
approaches typically focus on particular
admission policies, such as probabilistic
barring or age-dependent access. A broader
protocol framework is still needed to support
different device-local admission rules through
common signaling, while enabling the reader
to refine its guidance using subsequent
device feedback. This motivates a more selective and adaptive reader-triggered
access while retaining simple \ac{aiot} operation. We address this
through four contributions:
\begin{itemize}
    \item We propose a lightweight signaling mechanism that builds on \ac{3gpp} Release-19 by extending the paging message to carry \ac{acgi} and embedding optional device feedback within the \ac{d2r} data payload.

    \item We develop a generic distributed access-admission
    mechanism, where the reader broadcasts
    group-level guidance, devices independently
    determine their eligibility based on local
    conditions, and the reader adapts subsequent
    guidance using observed outcomes
    and optional feedback.

    \item We demonstrate the framework and its potential gains through a
    freshness-based device-monitoring case study,
    integrating reader-side bandit adaptation
    and benchmarking against conventional
    \ac{cbra}, device-side backoff, and
    probability-based guidance. We quantify
    the effects of traffic load, access resources,
    and paging-group configurations on report
    loss and delivery delay, showing the effectiveness of the proposed mechanism.

    \item We identify future research directions
    toward intelligent and resilient \ac{aiot}
    access, encompassing advanced reader-side
    adaptation, digital-twin-assisted protocols, and resilience-aware signaling
    and access procedures.
\end{itemize}

\section{Reader-Triggered Access in 3GPP A-IoT}
\label{sec:model}

\begin{figure}[t]
\centering
\includegraphics[width=\columnwidth]{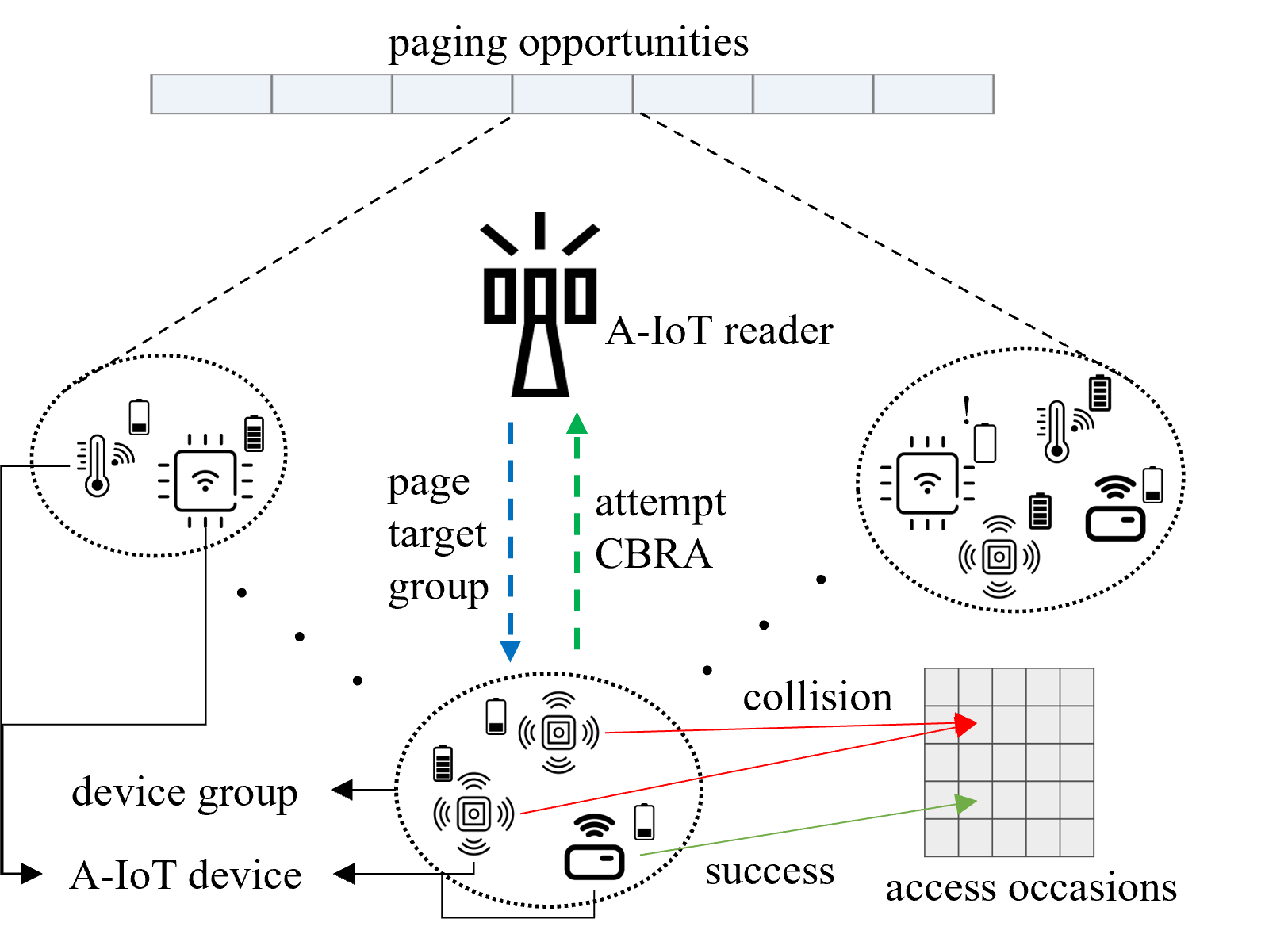}
\caption{Reader-triggered paging of \ac{aiot}
device groups and contention over shared
\acp{ao}.}
\label{fig:sysmod}
\end{figure}

As illustrated in Fig.~\ref{fig:sysmod},
the reader can address a selected group and
configure shared \acp{ao} for device responses~\cite{3gpp_38391}.
The addressed group may contain heterogeneous
A-IoT devices, such as sensing and tagging
devices with different operating conditions,
including energy-harvesting availability
and, where applicable, stored-energy levels. Under \ac{cbra}, each responding device
independently selects an \ac{ao} and transmits
its Access Random ID (MSG1). When multiple
devices select the same resource, their
transmissions may collide, leading to
unsuccessful access.

\begin{figure}[t]
\centering
\begin{tikzpicture}[
    >=Stealth,
    y=1.12cm,
    font=\footnotesize,
    life/.style={densely dashed,thin,black!60},
    msg/.style={
        align=center,fill=white,inner sep=1.4pt
    },
    exch/.style={-Stealth,semithick},
    optional/.style={
        -Stealth,densely dashed,thin
    },
    addition/.style={
        draw=black,
        densely dashed,
        fill=black!5,
        rounded corners=1pt,
        thin,
        align=center,
        inner sep=2pt,
        font=\footnotesize
    }
]
\def\xr{0}
\def\xd{6.9}

\node at (\xr,1.19) {A-IoT reader};
\node at (\xd,1.19) {A-IoT device};

\draw[life] (\xr,0.80) -- (\xr,-5.53);
\draw[life] (\xd,0.80) -- (\xd,-5.53);

\draw[exch] (\xr,0) -- (\xd,0)
    node[midway,above=2pt,msg]
    {A-IoT Paging (CBRA)\\
     $+$ \textbf{Proposed access guidance}};

\node[addition,text width=3.5cm]
    (local) at (4.8,-0.59)
    {Check paging and guidance\\
     Apply local admission rule};

\draw[black,densely dotted,thin]
    (local.east) -- (\xd,-0.59);

\draw[exch] (\xd,-1.62) -- (\xr,-1.62)
    node[midway,above=2pt,msg]
    {Access Random ID (MSG1)};

\draw[exch] (\xr,-2.52) -- (\xd,-2.52)
    node[midway,above=2pt,msg]
    {Random ID Response (MSG2)};

\draw[exch] (\xd,-3.48) -- (\xr,-3.48)
    node[midway,above=2pt,msg]
    {D2R upper-layer data transfer\\
     $+$ \textbf{Proposed optional feedback}};

\draw[optional] (\xr,-4.42) -- (\xd,-4.42)
    node[midway,above=2pt,msg]
    {Optional R2D data or NACK};

\node[addition,text width=4.2cm]
    (update) at (2.7,-5.18)
    {Update guidance\\
     for the next paging occasion};

\draw[black,densely dotted,thin]
    (\xr,-5.18) -- (update.west);

\begin{scope}[shift={(0,-6.17)},y=1cm]
\tikzset{
  protofield/.style={
      font=\scriptsize,align=center,inner sep=0pt
  },
  protolabel/.style={
      font=\footnotesize,anchor=west,inner sep=0pt
  }
}

\node[protolabel] at (0,0)
    {\textbf{R2D A-IoT Paging MAC PDU (CBRA)}};

\fill[black!7] (4.66,-0.96)
    rectangle (6.90,-0.24);

\draw[black!60,thin]
    (0,-0.96) rectangle (6.90,-0.24);

\foreach \x in {1.42,2.98,4.66}
    \draw[black!60,thin]
        (\x,-0.96)--(\x,-0.24);

\draw[black,densely dashed,thin]
    (4.66,-0.96) rectangle (6.90,-0.24);

\node[protofield,text width=1.28cm]
    at (0.71,-0.60)
    {Other\\147 bits};

\node[protofield,text width=1.42cm]
    at (2.20,-0.60)
    {Paging ID\\10 bits + var.};

\node[protofield,text width=1.52cm]
    at (3.82,-0.60)
    {Access cfg.\\23 bits};

\node[protofield,text width=2.12cm]
    at (5.78,-0.60)
    {Fill bits\\\textbf{Proposed AcGI}};

\node[
    font=\scriptsize,
    text=black!70,
    anchor=west,
    inner sep=0pt
] at (0,-1.20)
    {Fixed: 170 bits (no ID); 180 bits + ID bits (with ID)};

\node[protolabel] at (0,-1.67)
    {\textbf{D2R Upper Layer Data Transfer MAC PDU}};

\fill[black!7] (4.48,-2.62)
    rectangle (6.90,-1.90);

\draw[black!60,thin]
    (0,-2.62) rectangle (6.90,-1.90);

\foreach \x in {1.55,4.48}
    \draw[black!60,thin]
        (\x,-2.62)--(\x,-1.90);

\draw[black,densely dashed,thin]
    (4.48,-2.62) rectangle (6.90,-1.90);

\node[protofield,text width=1.38cm]
    at (0.775,-2.26)
    {MAC hdr.\\16 bits};

\node[protofield,text width=2.73cm]
    at (3.015,-2.26)
    {Upper-layer data\\variable};

\node[protofield,text width=2.17cm]
    at (5.69,-2.26)
    {\textbf{Proposed feedback}\\2 bytes};

\draw[black!55,thin]
    (1.57,-2.68)--(1.57,-2.77)--
    (6.88,-2.77)--(6.88,-2.68);

\node[
    font=\scriptsize,
    text=black!70,
    inner sep=0pt
] at (4.23,-2.98)
    {Data SDU: 0--123 bytes; padding omitted};

\end{scope}
\end{tikzpicture}
\caption{\ac{3gpp} reader-triggered
\ac{cbra} procedure and proposed signaling
extensions. Dashed boxes identify proposed
device-side and reader-side operations, while
the lower message structures illustrate potential
accommodation of access guidance within paging
Fill Bits and optional feedback within the D2R
Data SDU. The proposed elements are not
standardized \ac{3gpp} fields, and field widths
are schematic.}
\label{fig:guided_proc}
\end{figure}
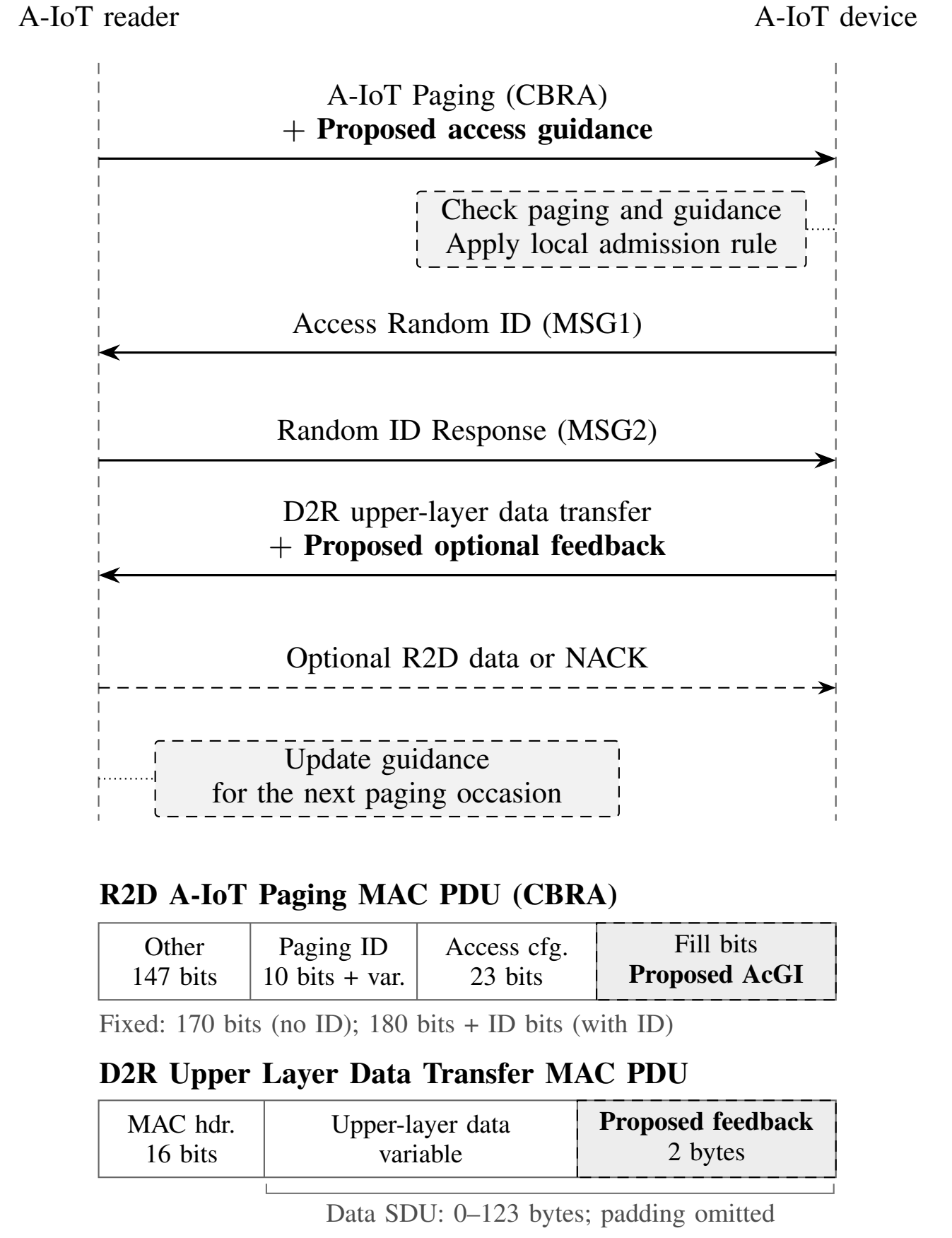

The corresponding signaling procedure is
illustrated by the black exchanges in
Fig.~\ref{fig:guided_proc}. Following successful
reception of MSG1, the reader can return
a Random ID Response (MSG2), allowing the
device to proceed with \ac{d2r} upper-layer
data transfer. Subsequent \ac{r2d} data
transfer or \ac{nack} feedback may be provided
when applicable~\cite{3gpp_38391}.
These exchanges establish the basic access
and data-delivery procedure without requiring
the reader to know individual device traffic
states beforehand.

The \ac{3gpp} paging message includes
several fields supporting this procedure.
In the \ac{cbra} format, these convey the
message type, transport block size, security
information, access type, transaction identifier,
and optional Paging ID. The latter identifies
the addressed devices, while its absence
allows paging of the entire population.
Additional fields specify the number of
\acp{ao}, transmission scheduling information,
and the response-monitoring configuration.
With the \ac{3gpp} security parameter present,
these fields occupy 170 bits without a Paging ID,
or 180 bits plus the variable-length Paging ID
when present. The message also includes optional
Fill Bits, which support byte alignment and are
explicitly identified as a possible location
for future extensions. 

The \ac{d2r} Upper Layer Data Transfer
\ac{mac} \ac{pdu} has a simpler structure.
Its 16-bit header identifies the message type,
indicates whether additional data remain,
and specifies the length of the data service
data unit (Data SDU). The Data SDU carries
up to 123 bytes of upper-layer information,
with segmentation supported when necessary. Optional
\ac{mac} padding accommodates unused
transport block capacity, while reserved
header bits have prescribed values.
The standardized message structures are
summarized in the lower part of
Fig.~\ref{fig:guided_proc} in black.

Although these procedures enable communication
with large device populations, the reader
has limited visibility into their individual
conditions. Successful receptions reveal
which transmissions were decoded, but
do not fully identify devices that remained
silent or experienced unsuccessful access,
nor the urgency of their undelivered reports.
Consequently, the reader cannot reliably
infer the complete access demand or the
relative importance of pending data from
its observations alone. This motivates
signaling mechanisms that can account for
device-local information without requiring
its acquisition before contention.

\section{Proposed Access Guidance and Reader Adaptation}
\label{sec:proposal}

Here, we extend reader-triggered access
with compact admission guidance and optional
post-access feedback. The central principle
is to distribute decision-making between
the reader and devices such that the reader provides
common access guidance, devices determine
their own eligibility, and subsequent
observations support adaptation at the reader.
The proposed additions to the Release-19
procedure are identified by dashed elements
and explicit labels in Fig.~\ref{fig:guided_proc}.

\subsection{Paging Guidance and Local Admission}
\label{sec:signaling}

Building on the \ac{3gpp} Release-19 provision for future
extensions through the paging Fill Bits, we
propose incorporating compact \ac{acgi} into
\ac{r2d} paging. This would enable devices to
make local admission decisions based on
reader-provided guidance and their individual
conditions, without requiring the reader to
acquire their states beforehand.

A candidate implementation introduces
a compact value or index within the
A-IoT Paging \ac{mac} \ac{pdu},
potentially referencing a predefined
parameter table. For example, a four-bit
index could select among 16 predefined
access probabilities, while a five-bit
index could specify a report-freshness
threshold. When multiple guidance policies
are supported, an additional indication
could identify the applicable rule. The reader and devices must share
the corresponding configuration and
interpretation.  As discussed earlier, the existing paging
Fill Bits offer a potential location for
such extensions. However, their available
capacity is variable, and under the current
\ac{3gpp} specification, devices treat them as
padding rather than interpreting their
contents as control information. Using
these bits for access guidance would
therefore require additional signaling
and processing rules. The exact encoding,
field allocation, capability indication,
and compatibility requirements therefore
remain subject to standardization.

Upon receiving a matching paging message,
a guidance-aware device evaluates the
advertised rule using its local information,
such as report age, priority, or access
history. Depending on the policy, admission may
involve checking whether a local condition
satisfies a broadcast threshold or making
a probabilistic access decision. An eligible
device proceeds with conventional \ac{cbra},
while a deferred device remains silent
for that paging opportunity. Importantly,
deferral does not necessarily imply report
deletion. Buffered reports remain subject
to application-specific replacement and
expiry rules. Guidance therefore regulates
participation in contention rather than
replacing the underlying access procedure.
Devices that do not support the extension
would require explicitly defined coexistence
and fallback behavior.

\subsection{Post-Access Feedback and Reader Adaptation}

To complement paging guidance, our proposed framework
introduces optional reporting of device-local
conditions following successful access.
Compact indicators may describe earlier
report losses, failed access attempts,
and the age of a delivered report.
Such information helps the reader assess
the consequences of its guidance decisions.
For example, a successfully reporting device
may reveal earlier report expiration or
unsuccessful access attempts that the reader
could not directly observe.

As illustrated in Fig.~\ref{fig:guided_proc},
these indicators could accompany the reading within the existing
\ac{d2r} Data SDU, provided that an
appropriate interface exposes them to
the reader-side controller. Alternatively,
a dedicated \ac{mac} extension could
provide reader-visible feedback, potentially
using otherwise unused padding capacity.
However, neither approach is currently
defined as a \ac{3gpp} access-guidance
feedback mechanism. The latter would
require explicit encoding and processing
rules, while the availability of padding
depends on the configured transport block
size. In either case, feedback overhead
and device energy constraints must be
considered.

Using successfully decoded reports and
any available feedback, the reader can
evaluate the performance of different
guidance parameters and refine subsequent
broadcasts. Adaptation may operate
independently for each paging group,
allowing different guidance selections
according to their observed traffic and
contention conditions. However, feedback
is inherently incomplete because devices that
remain silent or fail to deliver their
reports do not necessarily contribute
observations, potentially affecting the
accuracy of reader-side decisions.

This framework maintains a clear separation
of responsibilities. Devices retain their
local states and make admission decisions,
while the reader processes feedback,
evaluates performance, and updates guidance.
Any reward or cost calculations are confined
to the reader and need not be communicated
to devices. Practical implementations must
also define failure handling, how devices confirm delivery and reset counters, and behavior when guidance or feedback is unavailable. 


\subsection{Implementation Options and Trade-offs}

The proposed framework supports different levels
of signaling and control functionality, allowing
implementations to balance adaptability against
device complexity and communication overhead.
Several configurations can be considered:

\begin{itemize}
    \item \textbf{Mode-based guidance:}
    A one-bit indication could enable or disable
    a preconfigured admission rule, while a few
    additional bits could allow selection among
    multiple predefined policies. This approach
    requires minimal paging overhead and no
    additional device feedback. However, the
    available admission rules and their parameters
    must be known beforehand, limiting flexibility
    under changing network conditions.

\item \textbf{Parameterized guidance:}
The reader could broadcast a policy type
(e.g., probabilistic access or freshness-based
admission) together with an index selecting
a predefined parameter value, such as an
access probability or freshness threshold.
This allows the reader to adjust admission
decisions dynamically without requiring
additional device reports, although shared
parameter tables, capability signaling,
and sufficient paging capacity remain
important implementation considerations.

    \item \textbf{Feedback-assisted adaptation:}
    Optional D2R feedback could range from simple
    status indicators to richer representations
    of report losses, access failures, and
    information freshness. Such feedback could
    support more informed reader-side adaptation,
    but introduces additional payload, processing,
    and interface requirements. Its effectiveness
    also depends on the reliability and
    representativeness of successfully received
    reports.
\end{itemize}

These options illustrate how access guidance
could evolve from simple policy selection to
more adaptive and feedback-driven operation.
Their practical adoption would require
standardized signaling formats, clear
interoperability rules, and careful assessment
of the communication and energy costs.

\section{Case Study: Freshness-Guided \ac{aiot} \ac{cbra}}
\label{sec:implementation}

To illustrate the proposed framework, we consider an indoor
environmental monitoring scenario aligned with the 3GPP
\ac{aiot} sensor data collection use case~\cite{3gpp_22369}.
A reader collects measurements, e.g., temperature, humidity, and equipment-status, from \ac{aiot} devices deployed
in an indoor facility. Each device generates reports according to an
independent Poisson process, modeling asynchronous
sensor updates triggered by environmental changes
or monitoring events. This also allows for
evaluating different traffic loads without assuming
fixed reporting intervals. The reader pages one group
per simulation slot in round-robin order, with each slot
encompassing paging, contention, and data transfer.
Devices attempting access select a configured \ac{ao}
uniformly at random. This slot-based collision model
abstracts from physical-layer and energy-consumption details.
For this reporting-oriented evaluation, only paged devices
with valid buffered reports are modeled as access candidates,
isolating contention among devices seeking to deliver
sensor measurements.

Each device has a single-report buffer that retains
only the newest undelivered reading, with three states:
\begin{itemize}
    \item \textbf{Idle:} No report is buffered and a new arrival makes
    the device active.
    \item \textbf{Active:} A report is buffered but has not yet been
    attempted. An unsuccessful attempt makes it pending.
    \item \textbf{Pending:} A previously attempted report awaits
    another opportunity while unsuccessful retries leave it pending.
\end{itemize}
Successful delivery clears the buffer, while a new arrival replaces
any buffered report. If several readings arrive in a slot, only the
newest is retained. Reports also expire after their application-defined
lifetime. Thus, losses arise from replacement or expiry.

\subsection{Freshness-Based Admission and Compared Policies}
\label{subsec:score}

We implement \ac{acgi} as a threshold on report freshness. Each device
uses its report age, the group count, and the report lifetime to assign
the reading to an age band spanning one group-cycle length. For each
paged group, the reader broadcasts an index identifying the oldest
band eligible for access. The lowest index admits only the freshest
band, whereas the highest admits every unexpired report. Devices
apply this rule through a simple local comparison, without using
loss counters or previous access failures to determine eligibility.

For example, a 30-slot report lifetime and five paging groups yield
six age bands. Index two admits reports younger than 15 slots. The
bands describe elapsed group-cycle lengths, not how many paging
messages a report has actually encountered. Our evaluation assumes
that the reader and devices share the lifetime, group configuration,
slot-age convention, and index mapping, while heterogeneous lifetimes would
require additional configuration.

We compare four access policies:
\begin{itemize}

\item \textbf{Unguided CBRA:}
Following the 3GPP Release-19 reader-triggered
CBRA procedure~\cite{3gpp_38391}, paged devices
randomly select an available \ac{ao} and attempt
access without additional admission control.
Only devices holding valid buffered reports are
considered, excluding responses without data
to avoid unnecessarily degrading the baseline
and ensure a fair comparison.

    \item \textbf{Device-side backoff:}
    Each device initially attempts access with
    probability one, which is halved after each
    failed attempt for the same report, down to
    approximately 0.1\%. A new report resets the
    probability to one, and no guidance is broadcast.

    \item \textbf{Probability-only guidance:}
    The reader selects among 16 geometrically spaced
    access probabilities from 0.01 to one.
    Devices holding valid reports combine the
    broadcast admission probability with their
    local exponential backoff decisions.

    \item \textbf{Freshness-threshold guidance:}
    Devices first check whether their report-age
    band satisfies the broadcast threshold.
    Eligible devices then apply local exponential
    backoff before attempting access.
\end{itemize}
The two guided policies use the same reader-side
adaptation rule but differ in their candidate sets
and device-local admission decisions.

\subsection{Device Reports and Reader-Side Adaptation}
\label{subsec:feedback}

\begin{figure}[t]
\centering
\begin{tikzpicture}[
    >=Stealth,
    font=\footnotesize,
    flow/.style={-Stealth,semithick,black!75},
    guided/.style={-Stealth,semithick,densely dashed,draw=black},
    task/.style={
        draw=black!65,
        rounded corners=1.2pt,
        text width=2.62cm,
        minimum height=0.78cm,
        align=center,
        inner sep=3pt,
        fill=white
    },
    selected/.style={
        task,
        draw=black,
        densely dashed,
        fill=black!5
    },
    heading/.style={font=\footnotesize\bfseries}
]

\def\xl{1.50}
\def\xr{5.55}

\node[heading] at (\xl,0.94) {Device (local)};
\node[heading] at (\xr,0.94) {Reader};

\node[selected,minimum height=0.88cm] (agi) at (\xr,0)
    {\textbf{Proposed:} broadcast threshold\\in \acs{acgi} paging};

\node[task,minimum height=0.88cm] (match) at (\xl,0)
    {Match paging\\Read threshold};

\draw[guided] (agi.west) -- (match.east);

\node[task] (counters) at (\xl,-1.4)
    {Track report status,\\failed tries and age};

\node[task] (score) at (\xl,-2.9)
    {Determine report\\freshness level};

\node[task] (admit) at (\xl,-4.3)
    {Check eligibility\\Admit or defer};

\node[task,minimum height=0.99cm] (send) at (\xl,-6.4)
    {\textbf{If admitted}: perform CBRA\\
     \textbf{If delivered}: send reading\\
     and status feedback};

\draw[flow] (match.south) -- (counters.north);
\draw[flow] (counters.south) -- (score.north);
\draw[flow] (score.south) -- (admit.north);
\draw[flow] (admit.south) -- (send.north);

\node[task,minimum height=0.90cm] (decode) at (\xr,-1.13)
    {After successful D2R:\\read feedback};

\node[task] (loss) at (\xr,-2.26)
    {Estimate report-loss\\ penalty};

\node[task] (effort) at (\xr,-3.39)
    {Estimate access\\failure penalty};

\node[task] (age) at (\xr,-4.52)
    {Estimate report-age\\penalty};

\node[task] (reward) at (\xr,-5.65)
    {Compute round reward};

\node[selected,minimum height=0.96cm] (update) at (\xr,-6.85)
    {\textbf{Proposed:} update group estimate\\
     select next threshold};

\draw[flow] (decode.south) -- (loss.north);
\draw[flow] (loss.south) -- (effort.north);
\draw[flow] (effort.south) -- (age.north);
\draw[flow] (age.south) -- (reward.north);
\draw[flow] (reward.south) -- (update.north);

\draw[flow,rounded corners=3pt]
    (send.east) -- (3.48,-6.4)
    -- (3.48,-1.13) -- (decode.west);

\node[rotate=90,fill=white,inner sep=1.4pt]
    at (3.48,-4.00)
    {Successful D2R and feedback};

\draw[guided,rounded corners=4pt]
    (update.east) -- (7.51,-6.85)
    -- (7.51,0) -- (agi.east);

\node[rotate=90,fill=white,inner sep=1.4pt]
    at (7.51,-4.00) {Next paging};

\node[font=\footnotesize\bfseries]
    at (3.75,-7.75)
    {Proposed AcGI encoding (1 byte)};

\fill[black!10]
    (0.15,-8.65) rectangle (1.85,-8.05);
\fill[black!6]
    (1.85,-8.65) rectangle (5.50,-8.05);
\fill[black!3]
    (5.50,-8.65) rectangle (7.35,-8.05);

\draw[black!65,thin]
    (0.15,-8.65) rectangle (7.35,-8.05);

\foreach \x in {1.85,5.50}
    \draw[black!65,thin]
        (\x,-8.65) -- (\x,-8.05);

\node[font=\scriptsize,align=center]
    at (1.00,-8.35)
    {Guidance on/off\\1 bit};

\node[font=\scriptsize,align=center]
    at (3.675,-8.35)
    {Guidance index\\5 bits};

\node[font=\scriptsize,align=center]
    at (6.425,-8.35)
    {Spare\\2 bits};

\node[font=\footnotesize\bfseries]
    at (3.75,-9.19)
    {Proposed device feedback (2 bytes)};

\fill[black!10]
    (0.15,-10.09) rectangle (1.95,-9.49);
\fill[black!7]
    (1.95,-10.09) rectangle (3.75,-9.49);
\fill[black!5]
    (3.75,-10.09) rectangle (6.00,-9.49);
\fill[black!3]
    (6.00,-10.09) rectangle (7.35,-9.49);

\draw[black!65,thin]
    (0.15,-10.09) rectangle (7.35,-9.49);

\foreach \x in {1.95,3.75,6.00}
    \draw[black!65,thin]
        (\x,-10.09) -- (\x,-9.49);

\node[font=\scriptsize,align=center]
    at (1.05,-9.79)
    {Losses\\4 bits};

\node[font=\scriptsize,align=center]
    at (2.85,-9.79)
    {Failures\\4 bits};

\node[font=\scriptsize,align=center]
    at (4.875,-9.79)
    {Age\\5 bits};

\node[font=\scriptsize,align=center]
    at (6.675,-9.79)
    {Spare\\3 bits};

\end{tikzpicture}
\caption{Freshness-guided admission and reader-side adaptation.
Dashed elements denote the proposed guidance and feedback
mechanisms. The compact additional message structures are
illustrated at the bottom.}
\label{fig:guidance_reward}
\end{figure}

For the feedback process in Fig.~\ref{fig:guidance_reward}, each
device tracks reports replaced or expired and failed access attempts
since its last delivered report, together with the age of its current
reading. The first two counters accumulate across reports, while the age
refers to the reading eventually delivered. In the collision model,
an attempted access without a successful response increments the
failure counter. After successful delivery, the sensor reading and
these indicators are assumed to be available to the reader-side
controller through the upper-layer interface discussed in
Section~\ref{sec:proposal}.

The reader uses an $\epsilon$-greedy multi-armed
bandit~\cite{sutton2018reinforcement}, maintained
independently for each paging group. Each arm
corresponds to a predefined guidance setting
(i.e., an access probability or freshness threshold)
and maintains an estimate of its expected reward.

For each successfully delivered report, the reader
assigns one unit of reward, subtracting 0.25
for each previously lost report, 0.10 for each
earlier failed access attempt, and 0.50 times
the delivered report age normalized by its
lifetime. The loss and failure counts are
obtained from the device's post-access feedback. It sums the contributions and
divides by the number of configured \acp{ao},
while rounds without deliveries receive zero reward. Each delivered report's reward contribution is
floored at zero, preventing negative contributions
from making a round with deliveries score below
an empty round. The reward is computed from the feedback exactly as received,
i.e., from the saturated four-bit loss and failure counters and the report age.

Notably, failed attempts proxy access effort, not directly
measured energy consumption. Moreover, the reward is computed
only at the reader and is never transmitted to devices. After each paging round, only the selected arm's
reward estimate is updated, including with a zero
reward if nothing is delivered. At the group's next opportunity, the reader selects the highest-estimated candidate,
except with an exploration probability under which it selects a candidate at random, while
this probability decays from 0.1 towards 0.01 as the group accumulates observations.

The signaling requirements are illustrated in Fig.~\ref{fig:guidance_reward}. Specifically, 16
probability candidates need four information bits, while freshness
indices need three bits with five groups or five bits with one group
for a 30-slot lifetime. A common one-byte AcGI design could allocate
one bit to enable/disable the
preconfigured policy, five to the index, and two as spare bits. 
Feedback can use four bits each for report losses and failed
attempts, and five for report age: 13 bits packed into two bytes,
with three spare bits. Loss and failure counts exceeding 15 are
saturated. These are illustrative payload budgets rather than
standardized \ac{3gpp} field allocations. As stated earlier, allocating additional bits to
the policy indication could enable dynamic
switching among different admission strategies.

\subsection{Numerical Evaluation}\label{sec:results}

The default simulation configuration is motivated by the
\ac{3gpp} indoor environmental monitoring scenario,
which specifies a device density of 1.5 devices/m$^2$
and an end-to-end latency requirement of 20--30 s
~\cite{3gpp_22369}. We consider 380 devices,
corresponding to a nominal area of approximately
253 m$^2$, without explicitly modeling spatial
deployment. Assuming a one-second paging round,
the report lifetime is set to 30 slots, aligning
the application-level deadline with the upper end
of the specified latency range. The reader configures
16 \acp{ao} per round and a single paging group.
The average report-generation rate is 0.02 per device
per slot and is varied to assess the impact of traffic
load, as 3GPP does not prescribe a reporting interval
for this scenario.

Both guided schemes operate on top of device-side backoff.
Neither uses offline calibration, and each starts from its most
restrictive candidate index, with all reward estimates
initialized to zero. Learning runs online,
including during warm-up, and guidance adaptation is
performed independently per group.

Results are averaged over 50 independent realizations, paired across the evaluated
schemes, with 300 warm-up and 1,000 measurement slots per realization. Performance is
assessed through the report-loss ratio, defined as the fraction of reports generated
during the measurement period that are replaced or expire before successful delivery,
and the mean delivery delay of successfully delivered reports, measured in slots from
report generation. These metrics capture the trade-off between timely access and report
preservation, since restrictive admission may reduce delay while increasing losses.
Delay is omitted when the loss ratio reaches 95\%. In addition, the average
number of transmissions per delivered report can serve as a proxy for the effort/energy per successful
report.

\begin{figure}[!t]
\centering
\includegraphics[width=\columnwidth]{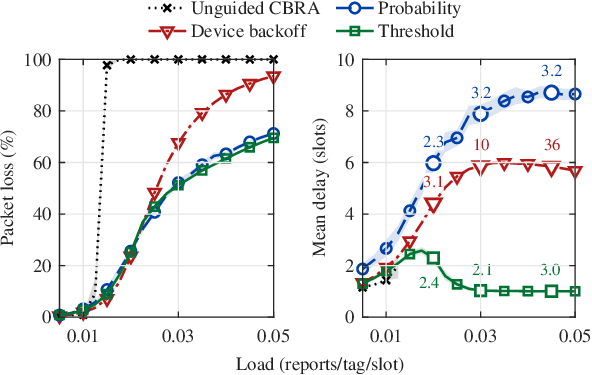}
\caption{(a) Report-loss ratio (left) and (b) delay of delivered reports (right) versus
the report-generation rate per device. Numbers next to open markers give the average
number of transmissions per delivered report and shaded areas show 95\% confidence
intervals.}
\label{fig:load}
\end{figure}

Fig.~\ref{fig:load} reveals three regimes as the report rate grows. Unguided \ac{cbra}
collapses as soon as traffic exceeds what the \acp{ao} can carry. At light and moderate
load, all adaptive schemes deliver most reports, and near the capacity transition, the two guided schemes are comparable. Under overload, threshold guidance loses the fewest reports, whereas device backoff degrades
the most. Threshold guidance also keeps the delay of delivered reports low and nearly
constant across all loads and needs the fewest transmissions per delivered report, while
the transmissions required by device backoff escalate as collisions accumulate.

\begin{figure}[!t]
\centering
\includegraphics[width=\columnwidth]{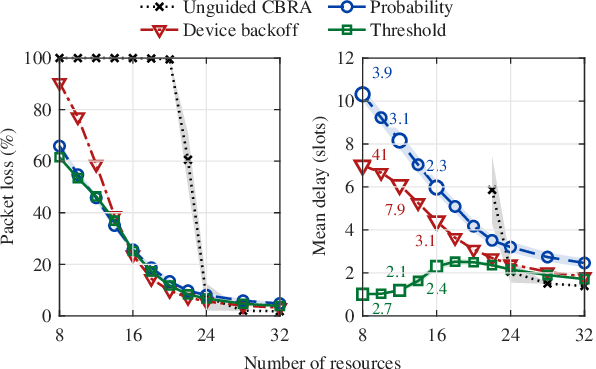}
\caption{(a) Report-loss ratio (left) and (b) delay of delivered reports (right) versus
the number of \acp{ao} per round. Numbers next to open markers give the average
number of transmissions per delivered report and shaded areas show 95\% confidence
intervals.}
\label{fig:resources}
\end{figure}

Fig.~\ref{fig:resources} shows the same regimes as a function of resources. When
\acp{ao} are scarce, \ac{cbra} loses almost all reports and device backoff degrades
the most among the adaptive schemes, whereas threshold guidance preserves the most
reports with the shortest delay and the fewest transmissions per delivered report. Near
the capacity transition, the guided schemes are comparable, and device backoff loses
slightly fewer reports. With sufficient \acp{ao}, all schemes perform well.

\begin{figure}[!t]
\centering
\includegraphics[width=\columnwidth]{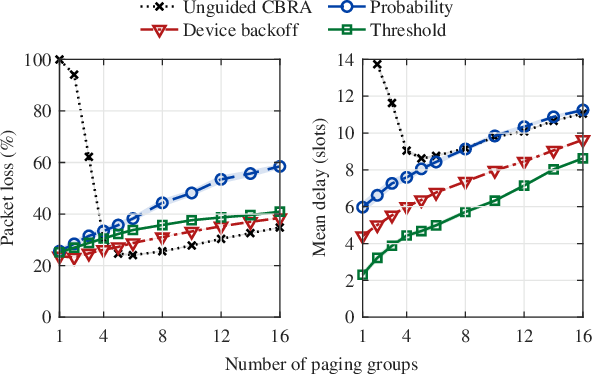}
\caption{(a) Report-loss ratio (left) and (b) delay of delivered reports (right) versus
the number of paging groups. Transmissions per delivered report are not annotated, as they are similar for all adaptive schemes
(1.4--3.1) across the curves.}
\label{fig:groups}
\end{figure}

Fig.~\ref{fig:groups} shows how the guidance adapts to the paging configuration, with
separate estimates per group. More groups mean fewer devices per paging occasion. With four or more paging groups, contention
decreases sufficiently for unguided CBRA
to recover, narrowing the performance gap
between the compared access schemes. However, excessive grouping also extends
the interval between paging opportunities,
increasing the likelihood of report replacement
or expiry in single-report buffers. This
leads to report loss eventually increasing
even as contention becomes less severe. Probability
guidance deteriorates as the number of paging groups increases, potentially reflecting
the reduced number of learning opportunities per group and its larger candidate set.
Freshness guidance requires fewer candidate thresholds under these configurations and
retains the shortest delay. However, this advantage narrows as grouping reduces
contention and all schemes need similar numbers of transmissions per delivered report.

Overall, freshness-threshold guidance performs best under severe contention, achieving
the lowest report loss together with the shortest delay and the fewest transmissions per
delivered report, while its outcomes vary far less across realizations than those of
probability guidance.

\section{Future Research Directions}
\label{sec:future}

Beyond the immediate benefits demonstrated in this work,
the proposed framework opens the door to a broader
evolution of \ac{aiot} access protocols, encompassing
several promising directions for future research
and standardization.

\subsection{Signaling and Feedback Design}

Further research and development should investigate the
encoding and placement of \ac{acgi} within existing
paging and control procedures, including optional
\ac{mac} extensions and device-capability indication.
Particular attention is needed to ensure coexistence
between guidance-aware and legacy devices, quantify
additional signaling overhead, and define how optional
access-related feedback can be made available to the
reader.

Beyond signaling design, feedback reliability is
equally important, since silent and unsuccessfully
contending devices cannot directly report their
conditions. Future designs should investigate compact
feedback representations, counter saturation, delayed
reporting, and mechanisms for confirming successful
delivery. Guidance policies must also remain effective
despite missing, quantized, or outdated observations.

\subsection{Context-Aware Admission and Intelligent Adaptation}

In addition to the freshness-based strategy,
device-local admission could also incorporate service
priority, energy availability, transmission history,
application-specific deadlines, or combinations of
these factors. Class-based admission, probabilistic
filtering, and multi-criteria eligibility rules offer
alternative implementations of the same signaling
framework. Their design should also address fairness
and starvation, ensuring that repeatedly deferred
devices are not systematically excluded from access.

Complementing these device-local decisions, more
advanced reader-side adaptation could improve guidance
selection under changing contention conditions.
Contextual bandits, reinforcement learning, and
traffic-prediction-assisted policies represent
promising approaches. Building on emerging developments
in \ac{ai}-assisted decision-making under
uncertainty~\cite{my_GenAI_IoT}, generative and
agentic \ac{ai} could further support joint
coordination of access guidance, paging-group
selection, and contention resources. However, their
benefits must be assessed against simpler controllers
in terms of performance, feedback requirements,
computational complexity, and convergence.

\subsection{Digital Twins, Validation, and Resilience}

Network digital twins could reproduce traffic,
paging, contention, and device behavior to evaluate
guidance policies and protocol extensions before
deployment. Beyond calibrating admission thresholds
and resource configurations, they could help
identify which additional signaling fields and
feedback mechanisms justify their overhead.
During operation, digital twins could also be
updated using observed statistics to anticipate
changing conditions and refine guidance decisions.

Establishing the practical value of these approaches
also requires evaluation under realistic operating
conditions, including correlated and event-triggered
reporting, heterogeneous traffic requirements,
intermittent energy harvesting, imperfect decoding,
capture effects, and multi-reader deployments.
Protocol-level and hardware-oriented validation
would help quantify practical gains and implementation
costs, while addressing modeling accuracy, incomplete
observability, and digital-twin mismatch.

An important extension is resilience to unexpected
disruptions, including traffic surges, device
unavailability, and changing service priorities.
Such conditions may require joint adaptation of
access guidance, paging configurations, and contention
resources, potentially beyond conventional
freshness-based admission.

\section{Conclusion}
\label{sec:conclusion}

This article presented a candidate \ac{acgi}
extension for reader-triggered \ac{3gpp}
\ac{aiot} \ac{cbra}, enabling devices to
make local admission decisions based on
reader-provided guidance while retaining
the underlying access procedure.
Through a freshness-based sensor-monitoring
case study, we demonstrated how compact
guidance and optional device feedback can
support adaptive contention management.
The numerical results showed improvements
in report loss and delivery delay,
particularly under congested conditions,
while highlighting the influence of traffic
load, access resources, and paging-group
configuration. More broadly, these findings motivate a shift toward
context-aware \ac{aiot} access, where reader-side
coordination and device-local decisions jointly
regulate contention within the existing
radio-access framework.

\bibliographystyle{IEEEtran}
\bibliography{references_full}

@techreport{3gpp_38391,
  author       = {{3GPP}},
  title        = {{NR}; Ambient {IoT} Medium Access Control ({MAC}) Protocol},
  institution  = {{3GPP}},
  type         = {Technical Specification},
  number       = {TS 38.391},
  year         = {2026},
  note         = {Release 19},
}

@techreport{3gpp_23369,
  author       = {{3GPP}},
  title        = {Architecture Support for Ambient Power-Enabled Internet of Things; Stage 2},
  institution  = {{3GPP}},
  type         = {Technical Specification},
  number       = {TS 23.369},
  year         = {2026},
  note         = {Release 19},
}

@misc{patent_group_suppression,
  author       = {G. Jo and H. Lee},
  title        = {Method and Apparatus of Preventing Data Transmissions for Ambient {IoT} in Wireless Communication System},
  howpublished = {International Patent Application WO 2025/198416 A1},
  year         = {2025},
  month        = sep,
}

@article{shen2025overview,
  author  = {Xiaodong Shen and others},
  title   = {State-of-the-Art Overview of {3GPP} Ambient {IoT} Radio Access Aspects},
  journal = {IEEE Commun. Stand. Mag.},
  volume  = {9},
  number  = {4},
  pages   = {215--223},
  year    = {2025},
  doi     = {10.1109/MCOMSTD.2025.3573815}
}

@ARTICLE{schoute1983dynamic,
  author={Schoute, F.},
  journal = {IEEE Trans. Commun.},
  title={Dynamic Frame Length {ALOHA}}, 
  year={1983},
  volume={31},
  number={4},
  pages={565-568},
  doi={10.1109/TCOM.1983.1095854}}

@ARTICLE{song2025twolevel,
  author={Song, Shilun and others},
  journal = {IEEE Commun. Lett.},
  title={Two-Level Estimation Enabled Online Congestion Control for Massive {IoT} Networks}, 
  year={2025},
  volume={29},
  number={8},
  pages={1968-1972},
  doi={10.1109/LCOMM.2025.3581943}}

@ARTICLE{song2026dynamic,
  author={Song, Shilun and others},
  journal = {IEEE J. Sel. Areas Commun.},
  title={Dynamic Multi-Channel Random Access Procedure for {6G}-Enabled Ambient {IoT}}, 
  year={2026},
  volume={44},
  number={},
  pages={2606-2621},
  doi={10.1109/JSAC.2025.3643814}}

@inproceedings{chen2020age,
  author    = {He Chen and others},
  title     = {Age-of-Information Dependent Random Access for Massive {IoT} Networks},
  booktitle = {IEEE INFOCOM Workshops},
  pages     = {930--935},
  year      = {2020},
  doi={10.1109/INFOCOMWKSHPS50562.2020.9162973}
}

@misc{ruizguirola2026adaptive,
  author        = {David E. Ruiz-Guirola and others},
  title         = {Adaptive Contention-based Random Access for Uplink Reporting in {3GPP} Ambient {IoT} Networks},
  year          = {2026},
  note          = {arXiv preprint arXiv:2605.04966},
  url           = {https://arxiv.org/abs/2605.04966}
}

@misc{my_GenAI_IoT,
      title={Toward Proactive {RF} Charging Scheduling: Generative {AI} for Decision Support}, 
      author={Amirhossein Azarbahram and others},
      year={2026},
      eprint={2606.10600},
      archivePrefix={arXiv},
      primaryClass={eess.SY},
      url={https://arxiv.org/abs/2606.10600}, 
}

@misc{patent_paging_threshold,
  author = {Zhibin Wu and others},
  title = {{Paging Techniques for Ambient IoT}},
  howpublished = {International Patent Application
    WO 2025/212854 A1},
  year = {2025},
  month = oct,
  url = {https://patents.google.com/patent/WO2025212854A1/en}
}

@misc{patent_congestion_control,
  author = {Zhikun Wu and others},
  title = {{Ambient Internet of Things Device
    Random Access Procedure}},
  howpublished = {International Patent Application
    WO 2025/231815 A1},
  year = {2025},
  month = nov,
  url = {https://patents.google.com/patent/WO2025231815A1/en}
}

@book{sutton2018reinforcement,
  author    = {Richard S. Sutton and Andrew G. Barto},
  title     = {Reinforcement Learning: An Introduction},
  edition   = {2},
  publisher = {MIT Press},
  year      = {2018}
}

@IEEEtranBSTCTL{BSTcontrol,
  CTLdash_repeated_names = "no"
}

@techreport{3gpp_22369,
  author      = {3GPP},
  title       = {{Service Requirements for Ambient Power-Enabled IoT}},
  institution  = {{3GPP}},
  type        = {Technical Specification},
  number      = {TS 22.369},
  version     = {19.3.0},
  year        = {2024},
  note        = {Release 19},
  url         = {https://portal.3gpp.org/desktopmodules/Specifications/SpecificationDetails.aspx?specificationId=4232}
}

@ARTICLE{David_context,
  author={Ruíz-Guirola, David E. and others},
  journal={IEEE Communications Magazine}, 
  title={Context-Awareness for Dependable Low-Power {IoT}}, 
  year={2026},
  volume={64},
  number={8},
  pages={86-92},
  doi={10.1109/MCOM.001.2500638}}

\section*{Biographies}

\noindent\textbf{Amirhossein Azarbahram} [S] is a postdoctoral researcher at the Centre for Wireless Communications (CWC), Oulu, focused on IoT, ISAC, and AI-native wireless systems. 

\noindent\textbf{David E. Ruiz-Guirola} [S'19, M'25] is a postdoctoral researcher focused on sustainable IoT, machine learning, and traffic prediction at the University of Oulu, Finland.

\noindent\textbf{Onel L\'opez} [S'17, M'20, SM'24] is an Associate Professor of wireless communications engineering, focused on sustainable IoT connectivity, at 6G Flagship, University of Oulu, Finland.

\end{document}